\documentclass[aps,prl,twocolumn]{revtex4}

\usepackage{graphicx}
\usepackage{amssymb}
\usepackage{hyperref}

\begin{document} 

\title{Gain-assisted slow to superluminal group velocity manipulation in nano-waveguides}

\author{Alexander A. Govyadinov}
\affiliation{Physics Department, Oregon State University, 301 Weniger Hall, Corvallis, OR 97331, USA}
\author{Viktor A. Podolskiy}
\email{vpodolsk@physics.oregonstate.edu}
\affiliation{Physics Department, Oregon State University, 301 Weniger Hall, Corvallis, OR 97331, USA}


\begin{abstract}
We study the energy propagation in subwavelength waveguides and demonstrate that the mechanism of material gain, previously suggested for loss compensation, is also a powerful tool to manipulate dispersion and propagation characteristics of electromagnetic pulses at the nanoscale. We show theoretically that the group velocity in {\it lossy} nano-waveguides can be controlled from slow to superluminal values by the material gain and waveguide geometry and develop an analytical description of the relevant physics. We utilize the developed formalism to show that gain-assisted dispersion management can be used to control the transition between ``photonic-funnel'' and ``photonic-compressor'' regimes in tapered nano-waveguides. The phenomenon of strong modulation of group velocity in subwavelength structures can be realized in waveguides with different geometries, and is present for both volume and surface-modes. 
\end{abstract}

\maketitle

Conventional optical fibers support propagating modes only when waveguide radius is sufficiently large\cite{Jackson}. In contrast to this behavior, plasmonic systems, anisotropy-based waveguides, nanoparticle chains, and optical coaxial cables \cite{BozhevolnyiPRL,podolskiyJMO,atwaterNature,keilmannMicron} support energy propagation even when the typical waveguide cross-section is much smaller than the wavelength. Unfortunately, since the majority of these nano-waveguides rely on plasmonic materials to confine the radiation beyond the diffraction limit, the propagation of nano-constrained radiation is often limited by material losses. While the emerging field of active plasmonics \cite{activePlasmonics} promises to overcome absorption limitations in nano-waveguides, full compensation of losses appears to be experimentally challenging\cite{MojahediCLEO}. 

In this Letter we focus on gain-assisted phenomena {\it beyond absorption compensation} and study the perspectives of controlling the dispersive properties of active nanoscale waveguides. We show that even relatively weak material gain, which is unable to compensate losses, is capable of producing large variations of the group velocity, bringing such exotic phenomena as slow ($0 < v_g \ll c$) and ultra-fast ($v_g < 0$) light\cite{SlowLight,Superluminal,LepeshkinSci} to the nanoscale domain. However, in contrast to diffraction-limited systems, where the group velocity is controlled solely by material dispersion, the energy propagation in nano-waveguides is also strongly affected by the waveguide geometry. We demonstrate that interplay between geometry- and material-controlled modal dispersion in tapered waveguides leads to the transition between {\it photonic compressor} regime, where the reduction of phase velocity is accompanied by the simultaneous reduction of group velocity \cite{joannopolous,stockmanPRL} and {\it photonic funnel} \cite{FunnelsPRB} regime where the product of phase and group velocities remains constant\cite{Jackson}. The developed formalism is illustrated on the examples of two fundamentally different nano-waveguides: a surface-mode-based plasmonic nanorod, and a volume-mode fiber with anisotropic core. Applications include \emph{nanosized tunable} delay lines, all-optical buffers and data synchronizers. 

\begin{figure}[b]
	\centerline{
		\includegraphics[width=8cm]{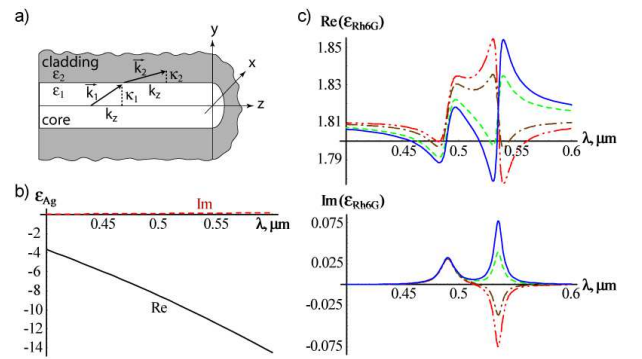}}
	\caption{(color online) (a) The schematic geometry of a multilayered waveguide; two-layer cylindrical system with axial symmetry is shown.  (b) Dielectric permittivity of $Ag$; real and imaginary parts are shown in solid and dashed respectively.  (c)	Real (top) and imaginary (bottom) parts of the dielectric permittivity of Rhodamine-6G model (see text). Solid blue, dashed green, dash-dotted brown and dash-dot-dotted red curves correspond to gain values $0\%$, $33\%$, $66\%$ and $100\%$ respectively. }
	\label{fig1}
\end{figure}

The geometry of a typical waveguide structure is schematically shown in Fig.~\ref{fig1}a. For simplicity, we assume that the system is homogeneous along the propagation direction ($z$ axis) \cite{footnoteWGHom}; the transverse structure of the waveguide is represented as a set of $N$ layers with different permittivities. Any wavepacket, propagating in such a structure can be described as a linear combination of {\it waveguide modes}, with each such mode having its own dispersion relation between propagation constant $k_z$ and frequency $\omega$. 

To find this relation, we first represent the field in $i$-th layer by a linear combination of ``incoming'' and ``outgoing'' (in radial direction) waves with the same $z$-dependence, but different structure across the layer (described by parameters $\kappa_i$)\cite{podolskiyJMO}. We then relate the fields in each layer to the fields in the neighboring layers through the boundary conditions, leading to the {\it waveguide matrix} $\mathbb{M}$ that relates the (complex) amplitudes of waves in all layers. Since a waveguide mode corresponds to a nontrivial solution of the system of algebraic equations described by $\mathbb{M}$, the modal dispersion equation can be found using:
\begin{eqnarray}
\label{eqDispTot}
\mathbb{D}(k_z, \omega,\vec{\kappa})=\det{\mathbb{M}}=0. 
\end{eqnarray}
where $\vec{\kappa}$ is a vector constructed of $\kappa_i$, which in turn are found from the wave- and layer-specific dispersion equations for individual waves:
\begin{eqnarray}
\label{eqDispWave}
D_i(k_z,\omega,\kappa_i)=0, 
\end{eqnarray}

Eqs.(\ref{eqDispTot},\ref{eqDispWave}) provide a complete description of electromagnetism in multi-layered waveguides. In particular, the group velocity is given by the following expression:
\begin{eqnarray}
\label{eqVgWaveguide}
	v_g=-\frac{\sum_{i}\frac{\partial \mathbb{D}}{\partial \kappa_i} \left( \frac{\partial D_i}{\partial \kappa_i} \right)^{-1} \frac{\partial D_i}{\partial k_z} - \frac{\partial \mathbb{D}}{\partial k_z} }
	{ \sum_{i}\frac{\partial \mathbb{D}}{\partial \kappa_i} \left( \frac{\partial D_i}{\partial \kappa_i} \right)^{-1} \frac{\partial D_i}{\partial \omega} -\frac{\partial \mathbb{D}}{\partial \omega} }.
\end{eqnarray}

The number of terms in the summations in Eq.(\ref{eqVgWaveguide}) (the length of $\vec{\kappa}$) depends on the waveguide symmetry and individual layer properties. Thus, when the system has axial symmetry (see Fig.~\ref{fig1}a), each layer with isotropic permittivity adds one $\kappa$ per ``axial number'' $m$, each layer with uniaxial anisotropy adds two such terms, etc. 

Eq.~(\ref{eqVgWaveguide}) can be used for an arbitrary waveguide geometry including planar, square, and circular systems, and can be applied to plasmonic, coaxial, and volume waveguide modes. To illustrate the developed formalism we consider two axially-symmetric cylindrical nano-waveguides: (I) a plasmonic nanorod in the dielectric material, and (II) a waveguide with anisotropic core, also known as photonic funnel \cite{FunnelsPRB}. 

In the case of the plasmonic waveguide the $z$ component of the $m$-th cylindrical wave inside and outside the metal nanorod can be represented by $I_m(\kappa_{1}r) e^{im\phi+i k_z z-i\omega t}$ and $K_m(\kappa_{2}r) e^{im\phi+i k_z z-i\omega t}$ respectively, with $I_m$ and $K_m$ being modified Bessel functions, and $\kappa_{1,2}^2=k_z^2-\epsilon_{1,2}\omega^2/c^2$. The boundary-matching technique described above yields the following set of dispersion equations for the first ($m=0$) $TM$ surface plasmon polariton (SPP) mode \cite{SPPs}:
\begin{eqnarray}
\label{sysPlasmon}
  \mathbb{D}^\textsc{spp}&=&\frac{\epsilon_1}{\epsilon_2}\frac{I_1(\kappa_1 R)}{\kappa_1 I_0(\kappa_1 R)}+\frac{K_1(\kappa_2 R)}{\kappa_2 K_0(\kappa_2 R)}
  \\
  D_{1,2}^{\rm SPP}&=&\epsilon_{1,2} \frac{\omega^2}{c^2}-k_z^2+\kappa_{1,2}^2. \nonumber
\end{eqnarray}
Direct substitution of Eqs.(\ref{sysPlasmon}) into Eq.(\ref{eqVgWaveguide}) yields:
\begin{eqnarray}
  v_g v_p &=& \left(\sum_{i=1}^2\frac{\partial \mathbb{D}^{\rm SPP}}{\partial \kappa_i} \frac{1}{\kappa_i}
\right)
\left[
  \sum_{i=1}^2\frac{\partial \mathbb{D}^{\rm SPP}}{\partial \kappa_i} \frac{1}{\kappa_i} 
  \left( \epsilon_i+\frac{\omega}{2} \frac{d \epsilon_i}{d \omega}\right)
    \right. 
\nonumber
\\
\label{eqVgVp}
&+& \left. \frac{c^2}{\omega} 
       \frac{\epsilon_1}{\epsilon_2} \frac{I_1(\kappa_1 R)}{\kappa_1 I_0(\kappa_1 R)} \left( \frac{1}{\epsilon_1} \frac{d\epsilon_1}{d\omega} - \frac{1}{\epsilon_2} \frac{d\epsilon_2}{d\omega} \right) \right]^{-1}
\end{eqnarray}
with $v_p$ being phase velocity of the mode:
\begin{equation}
	\label{Vp}
	v_p=\frac{\omega}{k_z}
\end{equation}

\begin{figure}[tb]
	\centerline{
	  \includegraphics[width=8.5cm]{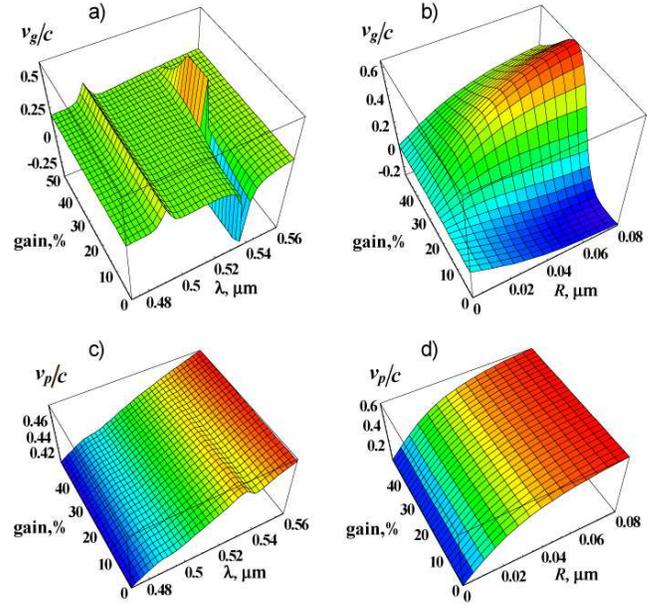}}
	\caption{(color online) Group and phase velocities of an SPP on metal nanorod in Rh6G methanol solution (see text) as functions of gain, frequency (a,c) and radius (b,d). The radial dependence is given at $\lambda = 534nm$, the wavelength one - at $R=35nm$.}
	\label{figPl}
\end{figure}

When the nanorod radius is much smaller than the free-space wavelength ($R \ll \lambda_0$), $\kappa_{1,2}$ is proportional to $1/R$. In this regime the phase velocity of SPP is proportional to the radius, the terms involving $\mathbb{D}^{\rm SPP}$ in Eq.~(\ref{eqVgVp}) are proportional to $R^3$, while the remaining term is proportional to $R$. 

The dramatic effect of material dispersion on the relationship between the phase and group velocities is now clearly seen. In the absence of material dispersion, $v_g v_p=\rm const$, and the decrease of phase velocity with the radius is accompanied by the {\it increase} of group velocity. This effect, introduced for anisotropic waveguides (see below)\cite{FunnelsPRB} is somewhat similar to the motion of uncompressible fluid through a funnel: the decrease of the cross-section leads to the increase of the local speed \cite{footnoteFluid}.

Any non-vanishing dispersion, however, dramatically changes this behavior and leads to the regime when both $v_p$ and $v_g$ simultaneously vanish. This regime, that originates from the material dispersion of plasmonic composites, has been shown to yield slow light in nanoscale plasmonic waveguides\cite{joannopolous} and in adiabatic plasmonic energy compressors\cite{stockmanPRL}. 

The crucial point of our work is that the dispersion of the \emph{dielectric core} can be used to control the group velocity of SPP by adjusting the term $\left( \frac{1}{\epsilon_1} \frac{d\epsilon_1}{d\omega} - \frac{1}{\epsilon_2} \frac{d\epsilon_2}{d\omega} \right)$ in Eq.~(\ref{eqVgVp}). This way the nanoplasmonic system switches between ``photonic funnel'' and ``photonic compressor'' regimes. Note that the group velocity can be independently controlled by either material dispersion or waveguide radius. It is therefore possible to build plasmonic systems with either ``fast'' or ``slow'' modes (see Fig.\ref{figPl}), or implement an adjustable gain mechanism to tune in between these two regimes. It is also possible to construct a tapered plasmonic fiber, similar to that described in\cite{stockmanPRL}, in which the packets would travel with superluminal speed at larger radii, and compress toward the small-radius apex.

To further demonstrate this behavior and show the control over the group velocity with the material gain we model the dispersive properties of a silver nanorod (described by Drude model, Fig.\ref{fig1}b):
\begin{equation}
\label{eqDrude}
\epsilon_1(\omega)=\epsilon^{Ag}=\epsilon_\infty^{Ag}-\frac{\omega_p^2}{\omega(\omega+i \gamma_p)}, 
\end{equation}
submerged into the 10\% solution (0.1-M) of  Rhodamine-6G in Methanol (Rh6G), which in the optical frequency range can be approximated by:
\begin{equation}
\label{eqLorentzian}
\epsilon_2(\omega)=\epsilon^{Rh6G}=\epsilon_{\infty}^{Rh6G}+\sum_{j=1}^2\frac{A_j}{\omega_{(0)_j}^2-\omega^2- i \gamma_j \omega}.
\end{equation}
In our calculations we use $\epsilon_\infty^{Ag}=5$, $\omega_p=46.26 \mu m^{-1}$, $\gamma_p=0.11 \mu m^{-1}$ for silver \cite{Palik} and $\epsilon^{Rh6G}_\infty=1.81$, $\gamma_1=0.4 \mu m^{-1}$, $\omega_{(0)_1}=12.82 \mu m^{-1}$, $\gamma_2=0.2 \mu m^{-1}$, $\omega_{(0)_2}=11.74 \mu m^{-1}$ for Rhodamine \cite{R6G}. We model the material gain by fixing $A_1=0.001$ and adjusting $A_2$ to gradually change the corresponding resonance strength. On the microscopic level, this process corresponds to a gradual increase of the population of the excited level of Rh6G with respect to the ground-state population, achievable, for example, by an external pump. The dependence of the dielectric permittivity of Rh6G on the material gain measured in percents of the excited state population is illustrated in Fig.~\ref{fig1}c. In the remaining of this work, we assume the regime with less than $50\%$ of population in the excited state, corresponding to $A_2>0$ ($A_2<0$ represents an inverted system). 

The group and phase velocities of SPP at the Ag-Rh6G interface are shown in Fig.~\ref{figPl}. As mentioned above, the gain level used in our work is relatively weak. Indeed, the phase velocity is almost unaffected by the material gain (see Fig.~\ref{figPl}c,d). Furthermore, the imaginary part of $\epsilon_2$, and of propagation constant $k_z^{\prime\prime}$ remain positive, indicating that the gain is insufficient to compensate SPP losses. Both superluminal ($v_g<0$) and slow ($0 < v_g\ll c$) light regimes can be identified in active nanoplasmonic structures. We note that while the SPP mode remains lossy, the total absorption in the fast- and slow-light regions is relatively small. The corresponding parameter $|k_z^{\prime\prime}/k_z|\lesssim 0.05$. 

When the $gain$ is close to $25\%$, the group velocity undergoes the transition from superluminal to slow light at a constant pumping level with varying nanorod radius. Similarly, at $R = 35 nm$, the superluminal to slow light transition happens as a function of pumping, providing almost unlimited control over the group velocity of SPPs at the nanoscale.

\begin{figure}[t]
	\centering
		\includegraphics[width=8cm]{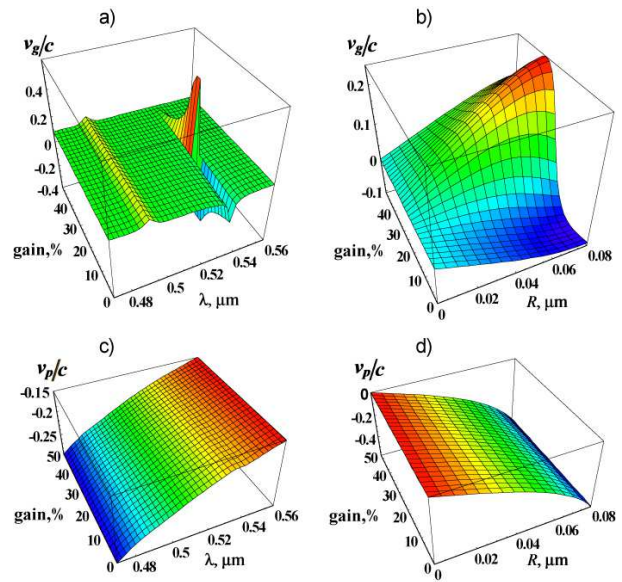}
	\caption{(color online) Group and phase velocities of the $TM_{01}$ mode in anisotropy-based cylindrical subdiffraction waveguide (see text) as functions of gain, frequency (a,c) and radius (b,d). Radial dependence is given at $\lambda = 534nm$, the wavelength one - at $R=35nm$. Note that the phase velocity is negative.}
	\label{figFun}
\end{figure}

To demonstrate universality of nanoscale group velocity modulation with respect to waveguide modes, we apply the developed formalism to another type of subdiffraction waveguide - the photonic funnel. These anisotropy-based systems support volume modes with either positive or negative refractive index \cite{layeredNIMs} and can be easily integrated with diffraction-limited fibers\cite{FunnelsPRB}. In the simplest case of a $TM_{01}$ mode in cylindrical geometry with uniaxial dielectric core with optical axis parallel to the direction of mode propagation and perfectly conducting metallic walls, Eqs.~(\ref{eqDispTot}-\ref{eqDispWave}) become:
\begin{eqnarray}
\label{sysFunnel}
  \mathbb{D}^\textsc{an}&=&J_0(\kappa_1 R)
  \\
  D^{\rm AN}_{1}&=&\frac{\omega^2}{c^2}-\frac{k_z^2}{\epsilon_{xy}}-\frac{\kappa_1^2}{\epsilon_{z}} \nonumber
\end{eqnarray}
where $\epsilon_{xy}$ and $\epsilon_{z}$ are the components of effective dielectric permittivity along and perpendicular to $z$-direction (see Fig.~\ref{fig1}a and ref. \cite{podolskiyJMO}) and $J_0$ is a Bessel function of the first kind. In the nanoscale limit ($R \ll \lambda_0$) Eq.~(\ref{eqVgWaveguide}) becomes:
\begin{equation}
\label{eqVgFun}
  v_g v_p = \frac{c^2}{\epsilon_{xy}} 
            \left[ 1 + \frac{c^2 \kappa_1^2}{2 \omega} 
              \left( \frac{1}{\epsilon_{xy}^2} \frac{d \epsilon_{xy}}{d \omega} + \frac{1}{\epsilon_z^2} \frac{d \epsilon_z}{d \omega} \right)
            \right]^{-1}.
\end{equation}
Similar to what has been shown earlier for SPP mode, this expression represents the interplay between the ``funnel limit'' $v_g v_p =  {c^2}/{\epsilon_{xy}} = \rm const$ and the ``compressor limit'' $v_g v_p\propto 1/\kappa_1^2 \propto R^2$.

Fig.~\ref{figFun} shows the behavior of the $TM_{01}$ mode in the waveguide with metamaterial core, consisting of alternating (in $z$ direction) layers of $Ag$ and Rh6G. In our computations we use Eqs.(\ref{eqDrude},\ref{eqLorentzian}) to calculate the dielectric constants of $Ag$ and Rh6G, accompanied by the effective-medium technique to obtain the effective permittivities of the nanolayer material (see Ref.\cite{podolskiyJMO} for details).

Note that the phase velocity (as well as the effective refractive index) of the metamaterial photonic funnel is negative\cite{FunnelsPRB}. Apart from the sign of $v_p$, the propagation of volume mode in metamaterial structure (Fig.~\ref{figFun}) is similar to that of an SPPs (Fig.~\ref{figPl}). Once again, one can effectively control the group velocity between ultra-fast ($v_g < 0$) and slow ($0 < v_g \ll c$) values {\it in lossy} ($|k_z^{\prime\prime}/k_z|\lesssim 0.08$) {\it nanoscale system} by tuning the material dispersion of the core or waveguide radius. 

To conclude, in this Letter we demonstrated the possibility of versatile (from slow to superluminal) group velocity modulation in sub-diffraction waveguides. The developed formalism can be directly utilized in different waveguide geometries (planar, square, etc.), different set of materials (polar dielectrics, semiconductors, quantum wells\cite{Qwells}, quantum dots \cite{Qdots}), and scaled for UV, optical, IR, THz and microwave spectral regions. Finally, we note that the strong modulation of group velocity can be also achieved in fully-active (amplifying) media, where it will be accompanied by loss-less or amplifying mode propagation\cite{activePlasmonics}.

The authors thank M.I. Stockman and R.Ziolkowski for fruitful discussions. This research has been partially supported by General Research Fund (Oregon State University).  

%This follows from the fact that Kramers-Kronig relations are symmetric under the complex-conjugation, i.e. the population inversion (see Fig. \ref{fig1}c) NOT SURE ABOUT THIS. While such inverted systems would consume more energy to operate, the lossless operation in conjunction with flexible velocity control will be even more beneficial to optical circuitry.

%Our analytical calculations show that by incorporating the material gain into a waveguide one can effectively control the group velocity at the nanoscale in an extremely broad range of speeds: from $v_g \ll c$ to $v_g < 0$. 


\begin{thebibliography}{}
\bibitem{footnoteWGHom} Although our formalism does not directly apply to nanoparticle chains since they rely on localized instead of propagating excitations, our results can be extended to these types of waveguides via incorporation of homogenization techniques (see for instance Ref.\cite{stroudPRB}).
\bibitem{footnoteFluid} Note however, that this analogy is not a complete one, since $v_g$ is proportional to the radius, not to crossection area.
\end{thebibliography}
\end{document}